# Incorporating Heightened Scrutiny into a Large HEP Software Project

*Philippe* Canal[1], *Chris* Green[1], *Kyle* Knoepfel[11], *Wim* Lavrijsen[2], *Adam* Lyon[1], *Marc* Paterno[1], *Saba* Sehrish[1], and *Beojan* Stanislaus[2]

[1]Fermi National Accelerator Laboratory, P.O. Box 500, Batavia, IL 60510, United States
[2]Lawrence Berkeley National Laboratory, 1 Cyclotron Road, Berkeley, CA 94720, United States

**Abstract.** In 2023, DUNE began re-evaluating the requirements of its data-processing framework, which led to commissioning a new design that would better fit neutrino physics than the existing reconstruction frameworks designed for collider physics. Due to the radical changes expected, significant multi-institutional effort has been directed toward the creation of the Phlex framework. In addition, the tight timelines in which to implement such a framework have invited scrutiny from various parties, including DUNE itself, the host laboratory Fermilab, and the US Department of Energy.

After briefly discussing the unique needs of neutrino physics, we will recount how the Phlex development team has approached the design and development of a framework in the context of this heightened scrutiny. This process began with a systems-engineering approach to formally manage the requirements DUNE has of its framework. What followed was, for the first time in the HEP community, a review of the framework's conceptual design by a panel of external framework experts before developers proceeded to the implementation. The implementation efforts have led to a series of prototypes that are being used to provide rapid feedback from Phlex users to framework developers.

We will discuss how each of these steps has resulted in a strong design that has the backing of the DUNE experiment, Fermilab, and the US Department of Energy.

## 1 DUNE's need for a more flexible framework

In HEP, the existing reconstruction frameworks are designed around the well-defined concept of an event. An event represents beam-crossing in collider-physics. The events are organized into fixed hierarchies of subruns and runs. The frameworks understand the hierarchies and how to process data that is organized in this manner. On the other hand, neutrino interaction signatures are very different in the form of detector readouts. While this approach of fixed hierarchy has been made to work for neutrino physics in the last decade or so, with the very large DUNE readouts, such an approach will not work. A typical readout for DUNE's far detector can be several GBs. The concept of an event does not match with what is needed for DUNE physics needs. Enforcing a fixed hierarchy to these large readouts is prone to memory issues.

[1] Corresponding author: knoepfel@fnal.gov

To support DUNE physics, a framework is needed that can allow the user-defined data groupings and have the ability to process according to the way data is organized. If a subsequent process needs a different data grouping, data can be re-grouped to facilitate that as well. These requirements on flexible data organization and processing led to the concept and design of Phlex. Phlex is a data processing framework that supports Parallel, Hierarchical, and Layered EXecution of data-processing algorithms being developed for DUNE as their core software framework. Phlex follows a data-flow oriented design, where data products are passed along the edges of a data-flow graph. Its design is highly motivated by concepts like high-order functions from functional programming.

The tight timelines of DUNE operations is shown in Figure 1. These timelines necessitated that a methodical process be adopted to develop a system to meet DUNE's data-processing needs. This includes defining requirements and establishing milestones leading to the design and implementation of the system. This led to a systems-engineering approach [1], where we gathered formal requirements from DUNE.

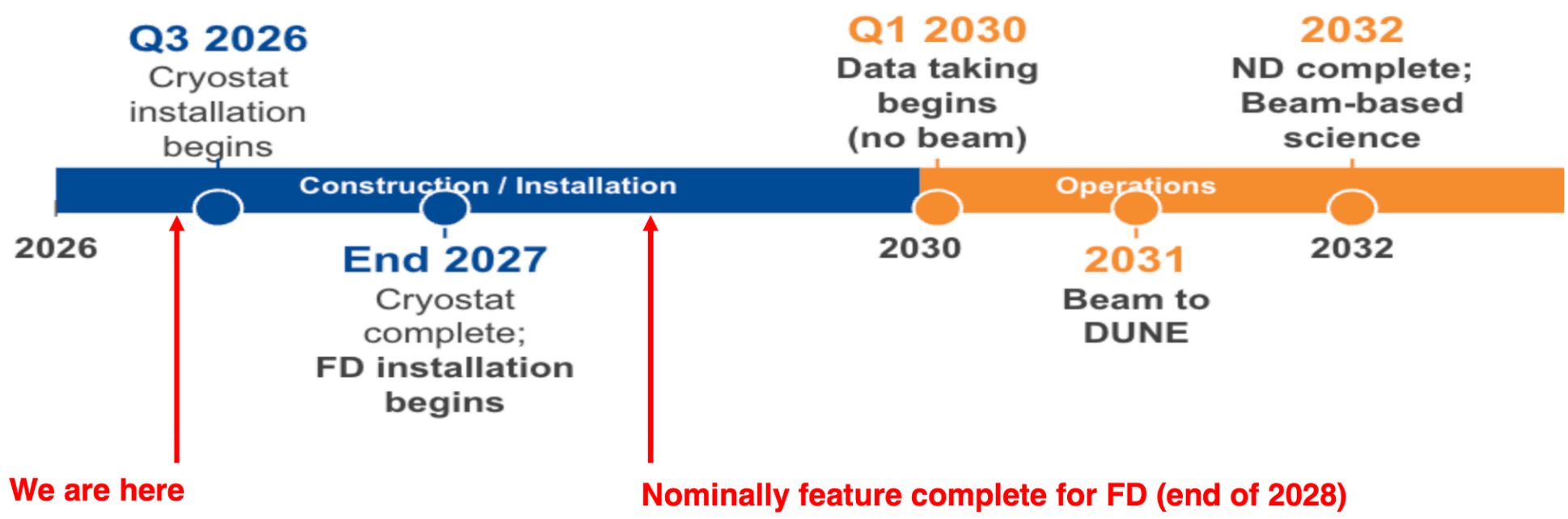

**Fig. 1.** Timeline of DUNE operations toward physics.

## 2 Systems engineering and framework requirements

In 2023 and 2024, DUNE commissioned a task force to identify requirements of a data-processing framework that would meet DUNE's physics needs. The requirements included the physics needs of its far detector, for which the *art* framework has been used, and the near detector, for which no formal data-processing framework has been used.

Throughout 2024 and 2025, the task force worked with framework developers at Fermilab and LBNL to refine the requirements to ensure that (a) each requirement focused on the need and not the solution; and (b) each requirement could be unambiguously verified. To facilitate this process, the concept of *user stories* were employed, where the user describes the physics case with a pattern like "With the framework, I want to X, so that I can Y." These user stories are then connected to one or more requirements that must be satisfied so that the user story can be achieved.

Managing such user stories and requirements requires a technology that can express relationships between user stories and requirements and track versions of those artifacts. Initial attempts of managing these artifacts using freely available software (such as Google sheets) quickly became untenable and a dedicated systems-engineering technology was pursued. The chosen technology is Jama Connect [2].

Jama Connect is a tool designed for systems-engineering needs. It supports three different kinds of requirements:

1. *Stakeholder* requirements, which are specified by DUNE and describe the highest-level behavior without describing implementations.
2. *System* requirements, which are middle-level requirements specified by the developers, describing subsystem requirements to meet the stakeholder requirements
3. *Component* requirements, which are lowest-level requirements, that (as of this writing) have not yet been used, but they describe the demands on the implementation to fulfill the system requirements.

Jama Connect supports user-story descriptions and expressible relationships between them and the requirements themselves (see Fig. 2). In addition, Jama Connect supports versioned user stories and requirements, and it allows recording the approval status of such items, enabling easy-to-understand snapshots of which items are settled and which ones need further work.

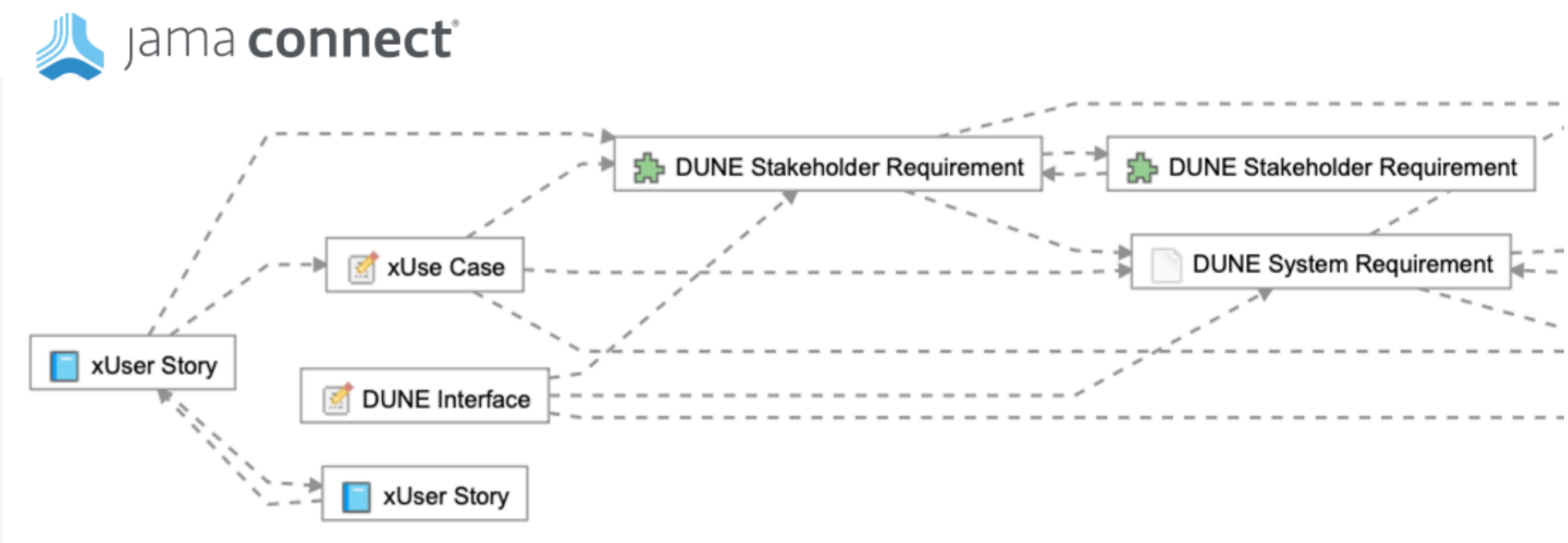

**Fig. 2.** Taxonomy of relationships among user stories, requirements, and other systems-engineering artifacts. The arrows point from higher-level items to lower-level ones. For example, user stories may depend on (a) other user stories, (b) stakeholder requirements, or (c) use cases, which are formal statements of framework behaviors under specific conditions.

Crucial to the systems-engineering process is the importance of clearly specifying requirements. To that end, Jama Connect provides an AI assistant that analyzes the text of a requirement and provides suggestions following EARS [3] and INCOSE [4] “best practices”. We found that following these suggestions clarified imprecise language that sometimes caused confusion. For example, one example of an imprecise requirement was:

> The framework MUST dynamically reduce or increase the number of concurrent modules processed based on available resources (e.g., CPU cores, memory, GPU, GPU memory, etc. and other co-processors), to improve efficiency.

After AI analysis, this requirement was refined to be:

> The framework shall dynamically schedule algorithms to execute efficiently according to the availability of each algorithm's required resources.

In the original, the list of example resources and the “etc” resulted in a requirement for which implementation compliance could not be verified. In addition, the tool flagged the use terms like “and” and “or”, which almost always indicate that multiple requirements are being expressed as one. The above replacement led to the development of 5 specific sub-requirements, each of which is verifiable.

The DUNE collaboration formally owns its stakeholder-level requirements, which the Phlex development team manages with Jama Connect. Although the development team may request clarifications (and even suggest them), the DUNE collaboration must approve additions, removals, and modifications. In contrast, the development team owns the system and component requirements. Jama Connect makes it possible for the DUNE collaboration to have a read-only view of all requirements so that DUNE can ensure that all requirements are consistent in satisfying their physics needs.

## 3 Reviewing the Conceptual Design

The use of software reviews for large projects is widespread in HEP. Historically these reviews are generally of already-implemented systems. For Phlex, both the developers and the DUNE experiment wanted a review of the concepts and high-level design of the framework before a full implementation was finished. Unlike the typical HEP software review, which is done on an implementation after it is complete or nearly complete, this would make it possible for the review to strongly influence the design and implementation of the framework. The review committee consisted of five scientists with extensive HENP framework development experience.

Conceptual design reports are commonly used in large projects in HEP to help structure the process of developing complex facilities and instruments. We chose to use this terminology for an analogous purpose in the design of Phlex. Our conceptual design document is intended to help the reader form the appropriate mental model for the system. The audience for the document is DUNE collaboration members, scientists from other collaborations who would like to determine whether Phlex can meet the needs of their experiment, reviewers, and representatives of funding agencies. The document concentrates on user-facing parts of the system: the data-flow graph, the several higher-order functions provided by Phlex, and what the registration and configuration of algorithms looks like. An important goal of the document is to allow the reader to understand *what* Phlex can do and how scientists would use it, rather than explaining the detailed mechanisms by which Phlex is implemented. The Phlex development team keeps the design report up-to-date as the design evolves and the current version is always available online [5].

This document was the primary source of information given to our external reviewers. To aid in their review process, the document was closely tied to the requirements we developed with DUNE. It contains an appendix with the full set of stakeholder requirements, and direct links to those requirements in Jama Connect.

The review was valuable to both DUNE and to the Phlex developers. The reviewers found: “We feel that Phlex has an ambitious design while still being based on firm technical footings. Its design confers clear advantages over existing reconstruction frameworks for neutrino experiments such as DUNE.” They declared that while the schedule was aggressive that the milestones could be met. To help with this, they identified some technical risks which, while manageable, could be mitigated. They also identified some of DUNE’s requirements that were imprecise or overly broad, and encouraged refinement of them.

# 4 Phlex development and continuous integration

Phlex is collaboratively developed using the Git version control system, and is hosted at https://github.com/framework-r-d/phlex. It has been released as Open Source software under the Apache 2.0 license, and all development is conducted in public.

Nevertheless the development model can best be described as a hybrid of the cathedral and bazaar models. The Phlex team meets on a weekly basis to discuss developments, and releases occur on a strict quarterly schedule. All pull requests must be reviewed by a member of the core development team before it can be merged. This also applies to pull requests from any of the core development team, which must be reviewed by a second core developer before merging.

The FORM I/O subsystem resides within the same repository as Phlex, and releases are made as part of Phlex. However the FORM team sets their own meeting schedule, and largely manages their own code.

The Phlex development workflow is heavily dependent on continuous integration and automated testing. The package has an extensive CI workflow built on GitHub Actions 4.5 thousand lines of YAML and 2 thousand lines of Python code. In addition to running all unit tests, the CI workflow also runs a clang-tidy linting pass and drives codecov to ensure test coverage is kept high. Testing is done on a variety of platforms and toolsets. Furthermore, pre-commit checks reduce the workload on the CI system and reduce pull request clutter. Finally, all C++ and Python code, and YAML workflow code are audited safety, security and code quality by CodeQL [6],

Artificial Intelligence is also exploited for code quality purposes, with every pull request reviewed by CodeRabbit [7]. AI-reviews have been found to be extremely useful for catching subtle bugs, while reducing the workload on human reviewers.

# 5 Prototype releases

The Phlex development team aligned the project's design requirements, as tracked in Jama, with specific DUNE milestones. This process established the functional requirements for a series of consecutive prototype releases. To solicit user feedback as early as possible, the team maintains an aggressive quarterly release schedule. This rapid turnover allows for continuous refinement of the feature sets of future framework releases.

All framework code and releases are publicly available. To help new users get started, a comprehensive set of installation instructions is provided, alongside a repository of working examples. While the current installation process relies on the Spack package manager, the long-term plan expects individual experiments to manage their own framework deployments. The current pre-1.0 releases are already functional enough for users to test the waters and in some cases start migrating their existing algorithms. Interested parties can track the full release schedule and view planned functionality at https://github.com/Framework-R-D/phlex/milestones.

# 6 Phlex's future

The recent CDR review adjusted how upcoming development steps are prioritized and highlighted areas where further guidance from DUNE is needed. Based on this review, some of the technical goals for the near future include: supporting calibration information as a first-class citizen, which requires a mechanism to associate data and calibration with intervals of validity; building a system to automatically and as needed translate data products between different memory layouts (e.g., from array of structures to structure of arrays and back for CPU/GPU interoperation) and programming languages (in particular Python-C++), which will have a registry for user-provided translations; and developing a design for persistence references and associations between data products.

In addition to technical goals, several collaborative goals guide near-term Phlex development. Primary among these are aiding DUNE's migration to Phlex and continuing the collaboration with Intel's oneTBB development team. This Intel collaboration is particularly timely; their interest in Phlex-specific use cases is driving support for features in oneTBB that will improve throughput for DUNE algorithms once implemented in Phlex. Furthermore, the framework's scope extends beyond DUNE. The development team is engaging with other experiments, including those outside neutrino research, that are interested in and currently testing Phlex. We expect that Phlex will be of benefit to other fields that rely on computing workflows to achieve scientific results.

## 7 Acknowledgments

This manuscript has been authored by Fermi Forward Discovery Group, LLC under Contract No. 89243024CSC000002 with the U.S. Department of Energy, Office of Science, Office of High Energy Physics.

This manuscript has been authored by an author at Lawrence Berkeley National Laboratory under Contract No. DE-AC02-05CH11231 with the U.S. Department of Energy. The U.S. Government retains, and the publisher, by accepting the article for publication, acknowledges, that the U.S. Government retains a non-exclusive, paid-up, irrevocable, world-wide license to publish or reproduce the published form of this manuscript, or allow others to do so, for U.S. Government purposes.

## 8 References

1. https://en.wikipedia.org/wiki/Systems_engineering
2. Jama Software, *Jama Connect* (Version 9.40.0). [Online]. Available: https://www.jamasoftware.com/.
3. A. Mavin, P. Wilkinson, A. Harwood, and M. Novak, "Easy Approach to Requirements Syntax (EARS)," in *2009 17th IEEE International Requirements Engineering Conference*, Atlanta, GA, USA, 2009, pp. 317–322, doi: 10.1109/RE.2009.9.
4. M. Ryan, L. Wheatcraft, R. Zinni, J. Dick, and K. Baksa, *Guide to Writing Requirements*, Doc. INCOSE-TP-2010-006-03.1, Ver. 3.1, San Diego, CA, USA: International Council on Systems Engineering (INCOSE), Apr. 2022.
5. Phlex Developer Team, *Phlex Framework Design,* [Online]. Available: https://framework-r-d.github.io/phlex-design/.
6. GitHub. (2026). *CodeQL: The libraries and queries that power security analysis* [Computer software]. GitHub repository. https://github.com/github/codeql
7. CodeRabbit, “CodeRabbit: AI-powered code reviews,” 2026. [Online]. Available: https://www.coderabbit.ai/.